# Non Destructive Testing

*G. Arnau Izquierdo*
CERN, Geneva, Switzerland

**Abstract**

The contribution introduces the principles, methods, and applications of non-destructive testing in the context of particle accelerators and related technologies is presented. Both surface inspection methods (visual testing, penetrant testing, magnetic particle testing, eddy current testing) and volumetric methods (radiographic testing, ultrasonic testing) are presented, with examples drawn from CERN projects. Emphasis is placed on the capabilities, limitations, and practical considerations of each technique, highlighting their role in ensuring quality and safety during fabrication, procurement, and in-service inspections. The contribution also addresses standards, codes, and personnel qualification schemes, underlining their regulatory impact in fields such as pressure equipment and industrial piping. Finally, the importance of early integration of NDT into project planning is stressed, not only for compliance but also as a proactive quality and efficiency measure.

## 1  Introduction

Non-destructive testing (NDT) refers to the practice of examining or obtaining information from materials, components, or assemblies without causing damage, thereby preserving their usability. This applies both to raw materials, which must remain suitable for manufacturing, and to finished components and assemblies, which must retain their functionality after inspection. NDT is also performed in service during maintenance programs or following the detection of anomalies.

From a historical perspective, one of the earliest industrial applications of NDT was the "oil and whiting" process, introduced in the late 1800s for detecting cracks in railroad components. The method involved applying oil to the surface of the component, cleaning it, and then applying a fine suspension of chalk in a solvent. Cracks became visible where the oil seeped out, staining the white coating. This principle was later refined into what is known today as dye penetrant testing (PT), which employs specially formulated penetrant liquids, developers, and sometimes ultraviolet light to detect very fine surface-breaking defects.

Contemporary NDT methods are grounded in the scientific principles of physics and chemistry. Several techniques are analogous to those used in medical diagnostics, including radiography, echography, and endoscopy. Table 1 lists the methods and abbreviations defined in the international standard *ISO 9712:2021, Non-destructive testing — Qualification and certification of NDT personnel* [1].

Table 1: Methods and abbreviated terms [1]

| NDT method | Abbreviated terms |
|---|---|
| Acoustic emission testing | AT |
| Eddy current testing | ET |
| Leak testing | LT |
| Magnetic testing | MT |
| Penetrant testing | PT |
| Radiographic testing | RT |
| Strain gauge testing | ST |
| Thermographic testing | TT |
| Ultrasonic testing | UT |
| Visual testing | VT |

## 2 Methods and principles

### 2.1 Surface methods

Those are methods aiming the detection of surface-breaking or near-surface defects.

#### 2.1.1 Visual Testing (VT)

Visual testing is the most basic and widely used NDT. It consists of either direct or indirect visual examination of a surface to detect visible imperfections such as cracks, corrosion, wear, or deformation.

Direct visual inspection can be performed with the naked eye, often aided by magnifying lenses, rulers, or lighting devices to improve visibility and accuracy. Indirect visual inspection employs optical instruments—including borescopes, fiberscopes, endoscopes, or video cameras—that enable access to confined or otherwise inaccessible areas. The choice of instrument depends on factors such as the size of the access, inspection distance, and required image quality, see Table 2.

Table 2: Optical instruments for indirect VT

| Type | Image transfer | Diameters | Lengths | Options |
|---|---|---|---|---|
| Borescopes | Lenses | 1-12 mm | 0.1-1 m | Fixed head |
| Fiberscopes | Optical fibre bundle | 0.5-10 mm | 0.1-3 m | Articulating head |
| Video-endoscopes | Image sensors (CMOS, CCD) | 2.5-12 mm | 1-30 m | Articulating head, exchangeable objectives |
| Inspection cameras | mage sensors (CMOS, CCD) | 20-200 mm | 3-1000 m | Rotating head, push rod, crawler |

High-quality professional video endoscopes are equipped with exchangeable tip optics, allowing targeted selection of the viewing direction (forward, radial, angled, or backward), depth of field (from 1 mm to infinity), and field of view (narrow, ~50°, to wide, ~120°). In some cases, they also enable dimensional measurements through stereoscopy or structured light projection, see Fig. 1.

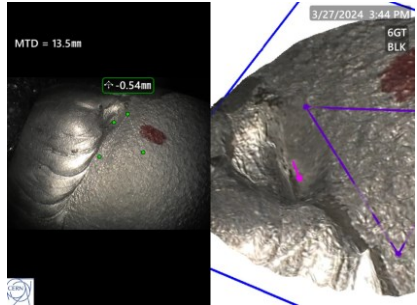

**Fig. 1:** Measurement of weld undercut depth using with a video endoscope with 3D capabilities.

Although VT is simple in concept, it is a critical step in nearly all inspection procedures. It is often the first method applied and can detect issues before more advanced techniques are used. However, its effectiveness depends heavily on the inspector's training, the accessibility of the surface, and the lighting and cleanliness of the part. For large components, detecting small defects can be time-consuming, and in such cases, more global methods may be considered.

### 2.1.2 *Penetrant Testing (PT)*

Penetrant Testing is a surface inspection technique used to detect open-to-surface discontinuities such as cracks, porosity, or laps. The method relies on the ability of a low-viscosity and low surface tension liquid, the penetrant, to seep into surface-breaking defects through capillary action. The visibility of the defect is enhanced by the spread out of the coloured dye on a contrasting white background or by the fluorescence of the dye in dark UV lighting.

The general procedure includes several steps: 1) surface preparation; 2) application of the penetrant and dwell time; 3) removal of excess penetrant; 4) application of the developer and 5) inspection to identify and interpret indications.

PT is applicable to both metallic and non-metallic materials, provided that the surfaces are non-porous. It is widely employed to detect very fine defects that may not be visible to the naked eye.

PT offers several practical benefits:

– It is simple and economical in terms of training requirements, equipment, and consumables.
– The method demonstrates high sensitivity to small discontinuities, such as cracks, porosity, or pitting.
– It is global, full large parts or batches of many small components can be tested at once.
– It is reliable regardless of the size of the part or the location of the discontinuity.

Despite its advantages, PT also presents several limitations:

– The handling and disposal of chemical products require strict safety measures.
– The method is limited to detecting surface-breaking discontinuities and cannot reveal subsurface or clogged defects.
– It is unsuitable for porous materials, which absorb penetrant.
– Certain components, such as those used in ultra-high vacuum (UHV) environments or those difficult to clean thoroughly, may be incompatible with the products used in PT.

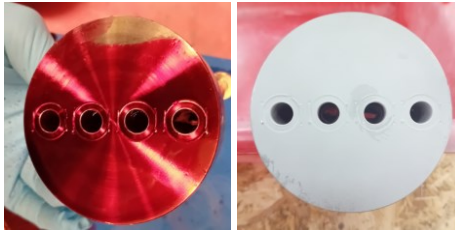

**Fig. 2:** PT performed on tube/plate welds in a cooling system for LHC 120 A current leads. After the application of the penetrant (left) and after the application of the developer (right)

### 2.1.3 *Magnetic Testing (MT)*

Magnetic Particle Testing, also referred to as magnetic testing, is employed to detect surface and near-surface discontinuities in ferromagnetic materials such as iron, nickel, cobalt, and certain alloys.

The principle of the method relies on magnetizing the part so that magnetic flux flows through it. When the flux encounters a discontinuity, such as a crack or inclusion, it is locally disturbed. If the discontinuity is close to the surface, the disturbance produces what is known as magnetic flux leakage. Fine ferrous particles, applied to the surface either as a dry powder or in a liquid suspension, are attracted to these leakage fields and form visible indications of the defect.

Several technique variants exist:

– Magnetization method: Direct magnetization (by passing electric current through the part) or indirect magnetization (by applying a magnetic field from an external source, such as a permanent magnet or a coil).

– Timing of inspection: Detection can be performed simultaneously with magnetization or based on residual magnetization.

– Detection medium: Magnetic particles applied as wet suspensions or dry powders, with black, colored, or fluorescent options.

– Equipment: Dedicated magnetic benches for larger components, or portable devices such as permanent magnets, electromagnets (yokes), and current generators.

MT is a quick, sensitive, and effective method for detecting cracks, laps, seams, and inclusions, particularly when the defect is oriented perpendicular to the applied magnetic field. However, its applicability is limited to ferromagnetic materials, and in many cases, surface preparation and demagnetization are required after inspection.

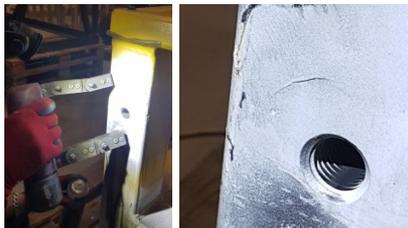

**Fig. 3:** MT of welds in a shipping frame for DUNE experiment cryostat. Magnetization (left) and detail of a crack indication (right)

### 2.1.4 *Eddy Current Testing (ET)*

Eddy Current Testing is an electromagnetic method primarily used for detecting surface and near-surface defects in conductive materials. It can also be used to measure coating thickness, assess material conductivity, and identify variations in heat treatment.

The principle of the technique relies on inducing circulating electrical currents—known as eddy currents—in the test piece. This is achieved by placing a probe coil carrying an alternating current in close proximity to the material. When eddy currents encounter a defect, such as a crack or corrosion pit, their flow is disrupted. This disturbance alters the coil impedance, producing measurable variations in reactance and resistance, which are used to detect and characterize flaws.

Some characteristics of ET are:

– Does not require a couplant; testing can be contactless and dry.
– Allow rapid scanning, making it suitable for high-throughput inspections in automated production lines.
– Provides high sensitivity to very small flaws in thin materials.
– Applicable only to conductive materials.
– Penetration depth decreases with increasing test frequency, limiting its effectiveness primarily to surface and near-surface defect detection.
– Signal interpretation requires significant expertise and training.

ET is widely used in aerospace for aircraft fuselage inspections, in the power industry for tubing and heat exchanger testing, and in manufacturing for the verification of metallic long products on production lines.

## 2.2 Volume methods

### 2.2.1 *Radiographic Testing (RT)*

Radiographic Testing is a volumetric inspection method that uses penetrating radiation — either X-rays or gamma rays — to create an image of the internal structure of a component. This image is captured on a radiographic film, a digital detector, or a phosphor imaging plate.

When the radiation passes through the material, it is attenuated according to the thickness and density of the part. Internal features such as cracks, porosity, inclusions, or lack of fusion in welds cause changes in attenuation, which appear as contrasts on the resulting image.

The main steps of RT include:

1. Selection of Parameters – Define the test arrangement by choosing the appropriate radiation source, detector type, exposure conditions (kV, mA, and time), and the number of views required to adequately cover the inspection area.

2. Safety Precautions – Implement all radiation protection measures. This includes the use of shielded rooms and interlocks, or, when in the field, clearly marking the exclusion zone, and carrying out radiation patrols to ensure the safety of personnel.

3. Positioning – Accurately place the source, film or detector, image quality indicator (IQI), and lead identification markers in accordance with the inspection plan.

4. Exposure – Perform the irradiation of the component under the defined parameters (time, kV, and mA), ensuring stable conditions for the duration of the exposure.

5. Image Development – Process the acquired radiographic image by chemical film development, digital scanning, or direct digital readout, depending on the equipment used.

6. Evaluation of the image quality – Verify image quality against applicable standards to validate or reject the image.

7. Evaluation of the object – If the image quality is acceptable, assess the inspected component to detect and characterize discontinuities. Document both image quality compliance and the conformity of the piece.

RT is highly effective for detecting volumetric defects and for documenting inspection results, but it requires strict safety measures to protect personnel from radiation exposure. It also demands careful selection of exposure parameters to ensure image quality.

Applications include weld inspection in pipelines, pressure vessels, and structural components, as well as the examination of castings and components. Figure 4 presents some examples at CERN.

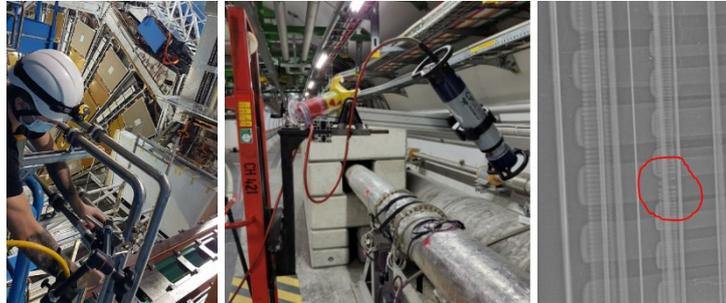

**Fig. 4:** Examples of RT in the field at CERN. Left, test of piping welds in the ATLAS experiment using a Se-75 gamma source. Middle and left, inspection with an X-ray generator of a warm module in the LHC and radiography showing a local relaxation of the sprig holding its contact fingers.

**Radiation Sources.** Radiographic testing can be carried out using a range of radiation sources. X-ray generators are available in both modular and portable configurations, with maximum voltages typically ranging from 75 to 600 kV. These devices feature focal spots between 0.4 and 5 mm and power ratings from 700 to 4500 W. Depending on the inspection requirements, they can produce directional, fan-shaped, or panoramic beams. Figure 5 from ISO 17636-2:2022 [2], presents the recommended voltage for X-ray generators as a function of penetrated thickness and material.

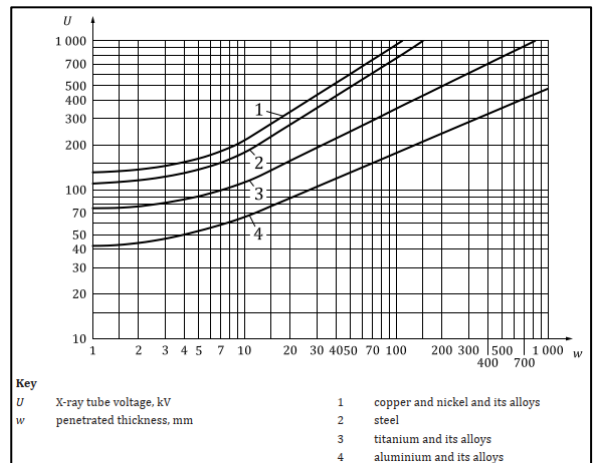

**Fig. 5:** Recommended voltage for X-ray generators as a function of penetrated thickness [2]

For higher energy applications, linear accelerators (linacs) may be employed. Alternatively, sealed radioactive isotopes are widely used, most commonly Selenium-75, Iridium-192, and Cobalt-60. These sources provide activities in the range of 0.5 to 8 TBq and source sizes between 1 and 3.5 mm, making them suitable for field applications and thick-section components.

**Detectors.** A variety of detectors are used to capture radiographic images. Traditional silver-based films remain common; they can be developed either manually or with automatic processors and are evaluated on a negatoscope using an intense light source. Computed radiography (CR) employs re-usable flexible phosphor imaging plates, where a latent image is formed and later revealed by scanning, producing a digital file expressed in grey levels. More recently, digital radiography (DR) uses rigid flat panel detector arrays that generate digital images directly, eliminating the development process. Each detection system offers a specific balance of characteristics in terms of resolution, sensitivity, flexibility, and handling.

**Radiographic Sensitivity.** The sensitivity of radiographic image is influenced by both contrast and definition. Contrast is affected by subject-related factors such as differences in material absorption, the energy of the primary radiation, and the contribution of scattered secondary radiation. Image definition is determined primarily by geometric conditions: the size of the radiation source, the source-to-detector distance (which controls geometric penumbra), the distance between the subject and the detector, the sharpness of thickness variations in the specimen, and any movement during exposure.

To verify image quality, inspectors use Image Quality Indicators (IQIs), typically consisting of wires or plates with drilled holes of decreasing size. These indicators are placed on the component and serve to confirm that the radiograph provides sufficient contrast and definition for proper evaluation.

**Test Arrangements for Welds.** For the inspection of welds, several radiographic arrangements are employed depending on the thickness and accessibility of the component. The single-wall technique involves exposing and recording through one wall only, providing a direct projection of the weld. In double-wall techniques, radiation passes through two walls before reaching the detector. The double-wall and double-image arrangement is commonly applied to small-diameter pipes, where both walls of the pipe and two images of the weld can be recorded in a single exposure.

The choice of arrangement depends on geometry, accessibility, and the level of detail required to detect potential defects. Figure 6 presents some test arrangements for but welds of curved objects, more arrangements are defined in ISO 17636-2:2022 [2].

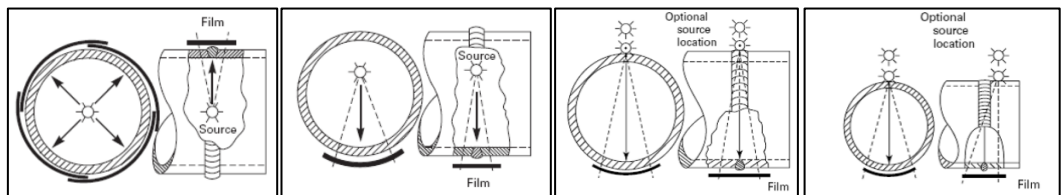

**Fig. 6:** Some single wall (left) and double wall (right) arrangements for RT of butt welds in curved objects

## 2.2.2  *Ultrasonic Testing (UT)*

Ultrasonic Testing is a volumetric inspection technique that uses high-frequency sound waves to detect internal flaws, measure thickness, and evaluate material properties. It is one of the most widely used NDT methods due to its sensitivity, depth of penetration, and versatility.

Ultrasonic Testing is based on the transmission of high-frequency sound pulses, typically between 0.5 and 20 MHz, into a material. The most common approach is the pulse–echo technique, in which sound waves are introduced into the specimen and echoes are detected either from internal flaws or from the back wall of the component. The results are displayed as a graph of signal amplitude versus time of flight, which can also be expressed as depth if the speed of sound in the material is known. Figure 7 presents a schematic diagram of the principle of the pulse-echo technique.

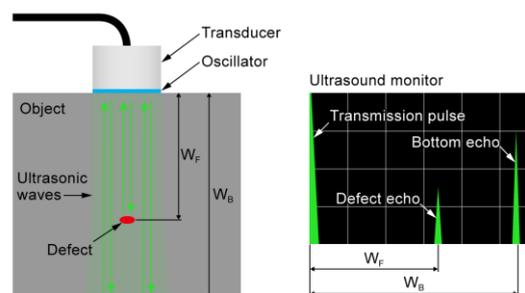

**Fig. 7:** Principle of the pulse-echo technique and the A-scan display of the signal

**Sound Field of a transducer.** The sound field generated by a transducer can be divided into two regions. In the near field the sound pressure undergoes alternating maxima and minima. This makes flaw evaluation by amplitude-based techniques difficult. Beyond this region lies the far field, where the sound pressure gradually decreases as the beam spreads and loses energy. The divergence of the beam is described by its spread angle, which affects resolution and coverage. Near field length and spread angle are function of transducer size, frequency and sound velocity in the object.

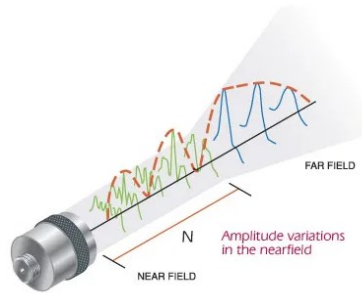

**Fig. 8:** The sound field generated in front of a straight-beam single-element probe

**Contact Probes.** Most inspections use contact probes, which require a couplant layer such as gel, oil, or water to ensure proper transmission of ultrasound into the material. These probes are typically used for manual inspections of parts with regular geometries and relatively smooth surfaces, whether flat or curved.

**Straight-Beam Techniques.** Single-element straight-beam probes are well suited for detecting flaws or back walls that are parallel to the inspection surface and are preferred for the penetration of thick sections. Dual-element, or TR probes, contain separate transmitter and receiver elements divided by a barrier to reduce crosstalk. These are particularly effective in thin sections or for detecting near-surface flaws where high resolution is required.

**Angle-Beam Techniques.** Angle-beam probes are mounted on wedges that refract the sound waves into the material at a predetermined angle. They typically use shear waves, although longitudinal waves can also be applied. Angle-beam inspection is especially suitable for welds and other components where flaws may be oriented at an angle to the surface.

**Immersion Transducer.** Immersion techniques are widely applied in mechanized or automated inspections. In this case, the component and the probe are separated by a column or bath of water, ensuring consistent coupling and reproducible results. Immersion transducers are particularly effective for inspecting joints with interfaces parallel to the entry face, such as brazed or diffusion-bonded assemblies. Large parts can be examined using probe manipulators or water jets, and the probes can be focused to enhance resolution.

**Phased Array Ultrasonic Testing (PAUT).** Phased Array techniques use probes made up of a matrix of small transducer elements, individually excited under computer control. By applying programmed delay laws, the combined sound beam can be steered or focused electronically. This enables the sequential generation of beams at different angles, producing angular scans (S-scans), or the application of dynamic focusing at varying depths.

**Representations of UT Data.** Ultrasonic results can be presented in several formats. The A-scan displays amplitude as a function of depth from a single static probe position. The B-scan represents amplitude as a function of lateral probe displacement (X-position versus depth), often with color coding for signal amplitude. The C-scan maps the inspection area in two dimensions (X versus Y) with color-coded amplitude or depth, producing a plan view of the component. The S-scan shows amplitude as a function of depth over a set of angles from a fixed probe position, providing a cross-sectional image through angular scanning.

Figures 9–11 present some examples of UT at CERN.

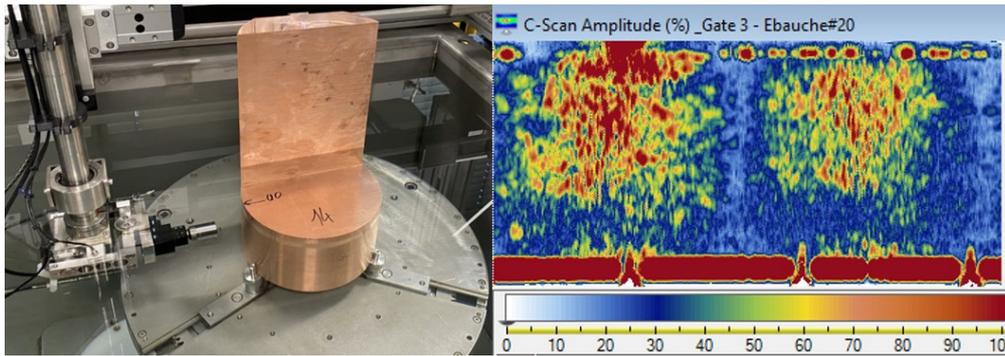
**Fig. 9:** Inspection by immersion of a copper blank for HL-LHC current leads showing the in depth of development of cracks in the cylindrical part

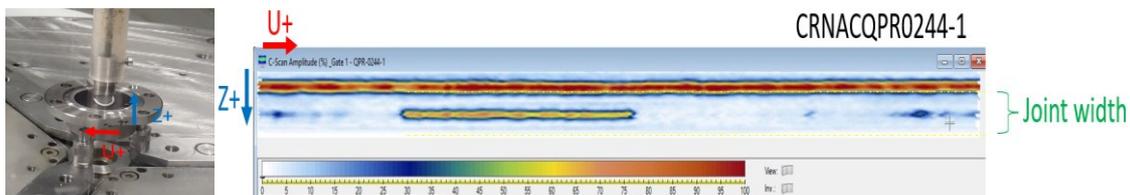
**Fig. 10:** Inspection by immersion of a brazed joint between a stainless-steel flange and a niobium sleeve for the cryostat of a superconducting RF cavity

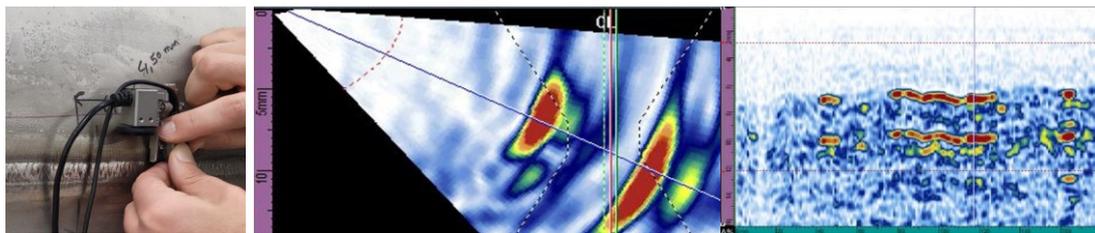
**Fig. 11:** Inspection by PAUT of a longitudinal weld in the cold mas of a superconducting magnet. Longitudinal scanning from a weld side (left), S-Scan on a position with a defect (middle) and B-scan along 160 mm (right).

# 3   Regulatory aspects

## 3.1   Codes, Standards, and Personnel Certifications

The use of NDT can carry regulatory implications, particularly in sectors such as pressure equipment, aerospace, nuclear power, … or ski lifts. In other cases, NDT is implemented as part of a manufacturer's or procurer's quality policy. Regulatory aspects apply both to the performance of inspections and to the qualification and certification of the personnel involved.

A broad body of codes and standards governs NDT practices. Standards are technical documents establishing requirements for products, processes, or methods. They are issued by organizations such as ISO, CEN (EN), ASTM, ASME, AWS, and NAS. Codes, by contrast, are legally binding once adopted by governmental bodies or referenced in contracts. They specify the minimum acceptable level of safety and typically include acceptance and rejection criteria for inspections. Examples include the EU Pressure Equipment Directive (PED) 2014/68/EU, the ASME Boiler and Pressure Vessel Code (BPVC), ASME B31.1 and B31.3 for piping, AWS D1.x Structural Welding Codes, and the French nuclear industry codes RCC-M and RSE-M.

NDT personnel qualification and certification is governed by several internationally recognized schemes. Under ISO 9712, examinations are carried out by accredited third-party national certification bodies, ensuring impartiality and harmonization across industries. In the United States, the ASNT scheme allows training and examinations to be conducted within employer-based certification systems.

For aerospace applications, employer-based frameworks such as EN 4179 and NAS 410 are used worldwide.

**3.2   Some examples of NDT in the Pressure Equipment Directive (PED) 2014/68/EU**

The PED is implemented through harmonized European standards such as *EN 13445 Unfired Pressure Vessels* and *EN 13480 Metallic Industrial Piping*. These standards include detailed provisions on the extent and type of NDT required. But NDT is not always imposed, let us see three examples.

*3.2.1   Open die forging blank in stainless steel X2CrNiMo18-14-3 as raw material for manufacturing a flange for a pressure vessel.*

In *EN 13445-2:2023 Unfired pressure vessels - Part 2: Materials* [3], the § 4.1.3 says that the materials shall be free from surface and internal defects which can impair their intended usability but does not require specific NDT. It redirects to *EN 10222-1:2021 Steel forgings for pressure purposes - Part 1: General requirements for open die forgings* [4], were one can see that MT, PT and UT are not part of the mandatory tests, they are optional. The requirements and conditions for NDT on that raw material may be agreed at the time of enquiry and order, and specific standards are only named in case NDT are to be performed. But nothing is imposed by default.

*3.2.2   Welded joints in a pressure vessel.*

In that case *EN 13445-5:2024 Unfired pressure vessels - Part 5: Inspection and Testing* [5] imposes a minimum extent of NDT defined as a function of weld type, material group, testing group, and NDT method; see Fig. 12. The quality requirement is generally Level C of EN ISO 5817, with stricter conditions applied for fatigue or creep service.

| Table 6.6.2-1 — Extent of non-destructive testing | | | | | | | | |
|---|---|---|---|---|---|---|---|---|
| TYPE OF WELD [a, p] | | | TESTING [b] | EXTENT FOR TESTING GROUP [o] | | | | |
| | | | | 1a | 1b | 2a [i] | 2b [i] | 3a | 3b |
| | | | | EXTENT FOR PARENT MATERIALS [l,m,n] | | | | |
| | | | | 1 to 10 | 1.1, 1.2, 8.1 | 8.2, 9.1, 9.2, 9.3, 10 | 1.1, 1.2, 8.1 | 8.2, 9.1, 9.2, 10 | 1.1, 1.2, 8.1 |
| Full penetration butt weld | 1 | Longitudinal joints | RT or UT | 100 % | 100 % | (100-10) % | (100-10) % | 25 % | 10 % |
| | | | MT or PT | 10 % | 10 % [d] | 10 % | 0 | 0 | 0 |
| | 2a | Circumferential joints on a shell, including circumferential joints between a shell and a non-hemispherical head | RT or UT | 25 % | 10 % | (25 -10) % | (10 - 5) % | 10 % | 5 % [c] |
| | | | MT or PT | 10 % | 10 % [d] | 10 % | 0 | 0 | 0 |
| | 2b | Circumferential joints on a shell, including circumferential joints between a shell and a non-hemispherical head, with backing strip [k] | RT or UT | NP | NA | NP | NA | NP | NA |
| | | | MT or PT | NP | 100 % | NP | 100 % | NP | 100 % |
| | 2c | Circumferential joggle joint, including circumferential joints between a shell and a non-hemispherical head [k] | RT or UT | NP | NA | NP | NA | NP | NA |
| | | | MT or PT | NP | 100 % | NP | 100 % | NP | 100 % |
| | 3a | Circumferential joints on a nozzle $d_i$ > 150 mm | RT or UT | 25 % | 10 % | (25 -10) % | (10 - 5) % | 10 % | 5 % [c] |

**Fig. 12:** Excerpt of EN 13445-5:2024, Table 6.6.2-1 [5]

*3.2.3   Welded joints in industrial piping.*

Similar provisions exist in the several parts of *EN 13480 Industrial Piping*. Welded joints must undergo a minimum extent of NDT depending on the weld type, material group, PED category, and NDT method; see Figs. 13–14.

**Table 8.2-1 — Extent of testing for circumferential, branch, fillet and seal welds**

| Material group [a] | Category | All welds | Circumferential welds | | | Branch welds | | | | | Socket/fillet welds | | Seal welds | |
|---|---|---|---|---|---|---|---|---|---|---|---|---|---|---|
| | | | Surface testing | | Volumetric testing [b] | Surface testing | | | Volumetric testing [b,k] | | Surface testing | | Surface testing | |
| | | VT % | $e_n$ mm | MT/PT [c] % | RT/UT % | Branch diameter | $e_n$ [h] mm | MT/PT [c] % | Branch diameter [i] | $e_n$ [h] mm | RT/UT % | $e_n$ mm | MT/PT % | $e_n$ mm | MT/PT % |
| 1.1, 1.2, 8.1 | I | 100 | 0 (5) [f,g] | | 5 (10) [g] | All | 0 (5) [f,g] | | All | | 0 | All | 0 | All | 0 |
| | II | | | | | | | | | | | | | | |
| | III | | | | 10 | | | 10 | > DN 100 | > 15 | 10 | | 10 | | 10 |
| 1.3, 1.4, 1.5, 2.1, 2.2, 4.1, 4.2, 5.1, 5.2, 8.2, 8.3, 9.1, 9.2, 9.3, 10.1, 10.2 | I | 100 | ≤ 30 | 5 | 10 | All [e] | | | All | | 0 | All [e] | 10 | All [e] | 5 |
| | | | > 30 | 10 | 10 | | | | | | | | | | |
| | II | | ≤ 30 | 5 | 10 | | | 10 (25) [g] | | | | | | | |
| | | | > 30 | 10 | 10 | | | | | | | | | | |
| | III | | ≤ 30 | 5 | 10 (25 [d]) [f,g] | All | | | > DN 100 | > 15 | 10 | All | 25 | All | 25 |
| | | | > 30 | 10 | 10 (25 [d]) [f,g] | | | | | | | | | | |
| 3.1, 3.2, 3.3, 5.3, 5.4, 6.1, 6.2, 6.3, 6.4, 7.1, 7.2 | I | 100 | ≤ 30 | 10 | 25 | 25 | | | > DN 100 | > 15 | 25 | 25 | 25 | 25 | 10 |
| | | | > 30 | 25 | 25 | | | | | | | | | | |
| | II | | ≤ 30 | 25 | 25 | All | | | | | | All | | All | |
| | | | > 30 | 25 | 25 (25 [d]) [f,g] | | | | | | | | | | |
| | III | | ≤ 30 | 100 | 25 (100) [f,g] | 100 | | | | | 100 | 100 | 100 | 100 | 100 |
| | | | > 30 | 100 | 25 (100 [d]) [f,g] | | | | | | | | | | |

[a] Material group, see CEN ISO/TR 15608.
[b] For the selection of the appropriate NDT-method for volumetric testing, see 8.4.4.3.
[c] See 8.4.4.2.
[d] Additional testing for transverse defects from weld surface (see EN ISO 17640:2010, testing level C).
[e] Only if PWHT has been carried out.
[f] Value in brackets applies to piping where creep or fatigue is the controlling factor in design.
[g] Value in brackets applies to piping with pneumatic pressure test with 1,1 times the maximum allowable pressure.
[h] $e_n$ is the nominal thickness of the branch pipe at the weld (see W3, W3.1 and W6 in EN 13480-4:2017, Figure 9.14.4-1 and Figure 9.14.4-2).
[i] For parts without DN designation $d_i$ > 120 mm may be used instead of DN > 100.
[k] Volumetric testing is required if both criteria (branch diameter and nominal thickness) are satisfied.

**Fig. 13:** Excerpt of EN 13480-5:224, Table 8.2-1 [6]

**Table 8.3-1 — Extent of NDT for longitudinal welds**

| Joint coefficient $z$ | VT % | MT or PT [a] % | RT or UT [b] % |
|---|---|---|---|
| $z ≤ 0,7$ | 100 | 0 | 0 |
| $0,7 < z ≤ 0,85$ | 100 | 10 | 10 |
| $0,85 < z ≤ 1,0$ | 100 | 100 | 100 |

[a] See 8.4.4.2
[b] See 8.4.4.3

**Fig. 14:** Excerpt of EN 13480-5:224, Table 8.3-1 [6]

As in EN 13445, the quality level is generally Level C of EN ISO 5817, with higher requirements for more demanding service conditions; see Fig. 15.

**Table 8.4.2-1 — Quality level according to EN ISO 5817:2014 depending on service conditions and test methods**

| Service conditions | Surface Imperfections and Imperfections in joint geometry | | Internal Imperfections |
|---|---|---|---|
| | Visual testing VT | Surface testing | Volumetric testing |
| Standard level | C | C | C |
| Fatigue | B | B | C |
| Creep | B | B | B |

**Fig. 15:** Excerpt of EN 13480-5:224, Table 8.4-2 [6]

# 4 Conclusions

A wide range of non-destructive testing methods and resources are available today, each with specific advantages, applications, and limitations. The diversity of techniques provides many possibilities but also requires careful consideration in order to select the most appropriate approach for each situation. It is therefore essential for engineers and project managers to be well informed about the capabilities and constraints of the different NDT methods.

NDT should be planned as an integral part of both fabrication and procurement processes. Its use should not be restricted to meeting only the minimum requirements imposed by codes or regulations. Instead, it should be regarded as a proactive quality decision, one that can facilitate production, improve efficiency, and reduce the risk of costly repairs or rework.

Finally, NDT must be integrated realistically into project schedules. Attempting to "squeeze" inspections into an already constrained timeline often compromises their effectiveness and may affect overall project quality. By considering NDT early, projects can benefit from smoother implementation and make inspections possible, easier, and more efficient.